\documentclass{osa-article}

\journal{osac}

\articletype{Research Article}

\usepackage{graphicx}
\usepackage{amsmath}
\usepackage[utf8]{inputenc}
\usepackage{float}
\usepackage{array}
\usepackage{makecell}
\usepackage{xcolor}
\usepackage{hyperref}
\usepackage{url}

\begin{document}

\title{Annealing by simulating the coherent  Ising machine}

\author{Egor S. Tiunov}
\address{Russian Quantum Center, 100 Novaya St., Skolkovo,
	Moscow 143025, Russia}
\address{Moscow Institute of Physics and Technology, Dolgoprudny 141700, Russia}
\author{Alexander E. Ulanov}
\address{Russian Quantum Center, 100 Novaya St., Skolkovo,
	Moscow 143025, Russia}
\author{A. I. Lvovsky}
\address{Russian Quantum Center, 100 Novaya St., Skolkovo,
	Moscow 143025, Russia}
\address{Department of Physics, University of Oxford, Oxford OX1 3PU, UK}
\address{P. N. Lebedev Physics Institute, Leninskiy prospect 53, Moscow 119991, Russia}
\email{Alex.Lvovsky@physics.ox.ac.uk}



\newcommand{\bra}[1]{\left\langle #1\right|}
\newcommand{\ket}[1]{\left| #1\right\rangle}
\newcommand{\braket}[2]{\left\langle
#1\vphantom{#2}\right|\left.#2\vphantom{#1}\right\rangle}
\newcommand{\ketbra}[2]{\left| #1\right\rangle\!\left\langle#2\right|}
\newcommand{\avg}[1]{\left\langle #1\right\rangle}
\newcommand{\be}[0]{\begin{equation}}
\newcommand{\ee}[0]{\end{equation}}
\newcommand{\intinf}[0]{\int_{-\infty}^{+\infty}}
\newcommand{\ada}[0]{\hat a^\dagger\hat a}
\newcommand{\ad}[0]{\hat a^\dagger}
\newcommand{\ah}[0]{\hat a}
\newcommand{\hv}[1]{\hat{\vec{#1}}}
\newcommand{\tr}[0]{{\rm Tr}}
\newcommand{\re}[0]{{\rm Re}\,}
\newcommand{\im}[0]{{\rm Im}\,}
\newcommand{\lra}\leftrightarrow
\newcommand{\eeqref}[1]{Eq.~(\ref{#1})}
\newcommand{\braketop}[3]{\left\langle
#1\vphantom{#2#3}\right|\left.#2\vphantom{#1#3}\right|\left.#3\vphantom{#1#2}\right\rangle}
\newcommand{\expec}[1]{\left\langle #1\right\rangle}
\newcommand{\zero}[0]{\ket{{\rm zero}}}
\newcommand{\adj}[1]{{\rm{Adjoint}}\left({#1}\right)}
\newcommand{\mat}[2]{\left(\begin{array}{#1} #2 \end{array}\right)}
\newcommand{\de}[0] {{\rm d}}
\newcommand{\os}[1] {\overset{\eqref{#1}}{=}}
\newcommand{\nna}[0] {\nonumber \\}
\newcommand{\nnb
}[0] {\\ \nonumber }
\newcommand{\iea}[0]{{\it et al.~}}
\newcommand{\ieac}[0]{{\it et al., }}
\newcommand{\sq}[0]{\ket{{\rm sq}_R}}
\newcommand{\tmsv}[0]{\ket{{\rm TMSV}_R}} 



\begin{abstract}
	The coherent Ising machine (CIM) enables efficient sampling of low-lying energy states of the Ising Hamiltonian with all-to-all connectivity by encoding the spins in the amplitudes of pulsed modes in an optical parametric oscillator (OPO).  The interaction between the pulses is realized by means of measurement-based optoelectronic feedforward which enhances the gain for lower-energy spin configurations.  We present an efficient method of simulating the CIM on a classical computer  that outperforms the CIM itself as well as the noisy mean-field annealer in terms of both the quality of the samples and the computational speed.  It is furthermore advantageous with respect to the CIM in that it can handle Ising Hamiltonians with arbitrary real-valued node coupling strengths. 
	These results illuminate the nature of the faster performance exhibited by the CIM and may give rise to a new class of quantum-inspired algorithms of classical annealing  that can successfully compete with existing methods.
\end{abstract}

\date{\today}



\paragraph{Introduction.}
The Ising model, originally developed for the description of phase transitions in magnetic materials \cite{baxter2014}, is a pillar of many branches in science. Its applications range from  quantum field theory \cite{IsingInQFT} and quantum gravity \cite{IsingInQGrav} to economics \cite{IsingInEcon} and machine learning \cite{Biamonte2017}. The model is defined by the Hamiltonian
\begin{equation}\label{IsingHam}
H = -\frac{1}{2}\sum_{i,j}J_{ij}\sigma_i \sigma_j 
\end{equation} 
where $J_{ij} = J_{ji}$, $J_{ii}=0$. The spin variable $\sigma_i$ can be 
a quantum operator or a number, continuous or discrete. In this paper we are interested in the classical discrete case, $\sigma_i=\pm1$. The goal is to find the configurations of spins which realize or approach the global minimum of the Ising Hamiltonian (\ref{IsingHam}). This is an NP-hard problem, meaning that the number of operations needed to find the exact solution grows exponentially with the size of the spin system \cite{Barahona1982}.  However, there exists a variety of classical algorithms to find an approximate solution \cite{Kirkpatrick671, MeanField, Smith1999, BLS}. 

Recently, the Ising problem has been approached using noisy intermediate-scale quantum (NISQ) devices. One example is the D-Wave quantum annealer based on superconducting qubits. However, it has a shortcoming associated with low connectivity: each physical qubit in the D-Wave chimera graph architecture is connected to only eight others. Therefore each logical spin must be embedded in multiple physical qubits to model the fully connected Hamiltonian (\ref{IsingHam}), resulting in a significant overhead \cite{DwaveMapping}. 

Another example of NISQ technology is the coherent Ising machine (CIM) --- an optoelectronic device developed in 2016 \cite{Inagaki2016,Mcmahon2016}. Experiments seem to demonstrate that CIM outperforms not only classical simulated annealing algorithms \cite{Haribara2017}, but also the D-Wave annealer \cite{CIMvsDwave}; however, the latter result is being disputed by the D-Wave team \cite{CIMvsDwaveComment}.

It has been argued that the fast performance of the CIM is of quantum nature, arising thanks to the nonclassical nature of optical squeezing \cite{Yamamoto2017}. This view, however, challenges the universally accepted belief that, in order to exhibit quantum speedup, a computational device would require large-scale entanglement and thoroughly minimized interaction with the environment, as CIM does not possess either of these features. A reasonable alternative hypothesis would be to ascribe the speedup to the analog optical processing that takes place in the CIM. Indeed, it is known that such processing, even in the absence of quantum effects, can be used to parallelize computational operations \cite{Caulfield2010}.  

To address these questions, a number of approaches to computer simulation of the CIM and related physics have been developed \cite{CIMvsDwave,Wang2013,Clements2017,King2018,Kalinin2018,leleu2018destabilization}. In particular, a recent study \cite{King2018} made an explicit comparison of a mean-field digital annealer and observed advantage of their algorithm with respect to the CIM in terms of both the annealing time and quality of samples obtained. 
In the present work, we develop a new simulator, which we call SimCIM because it is largely based on the actual physics of the CIM. We compare its performance with the \emph{bona fide} CIM and the annealer of Ref.~\cite{King2018} and find the results of this comparison to be generally in favor of our algorithm.

\begin{figure}
\centerline{	\includegraphics[width=0.6\linewidth]{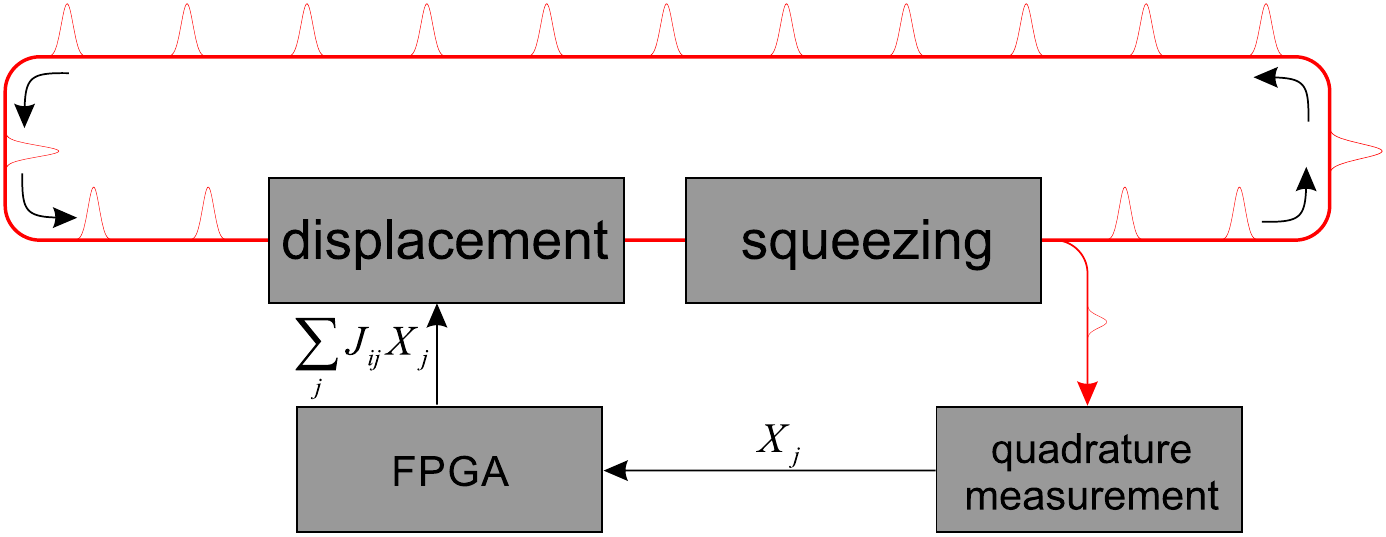}}
	\caption{CIM setup. Each pulse undergoes optical squeezing, linear and non-linear loss, as well as displacement. FPGA: field-programmable gate array.}\label{CIMFig}
\end{figure}

\paragraph{The coherent Ising machine.}
The primary element of the CIM architecture (Fig.~\ref{CIMFig}) is an optical parametric oscillator (OPO) made up of a fiber loop and a degenerate parametric amplifier (single-mode squeezer). The OPO is pumped above threshold in a pulsed manner. The length of the loop is matched  to an integer number of intervals between pump pulses, which results in well-identified pulsed modes circulating inside the OPO. After multiple roundtrips, these modes are amplified from the vacuum state to a microscopic amplitude. Due to the phase-sensitive nature of the squeezer, the phases of the resulting modes tend to either $\varphi_i=0$ or $\pi$. 

This symmetry is broken by a measurement-based pairwise interaction between pulses, which is arranged as follows. Upon exiting the amplifier, a part of the optical energy is deflected to a homodyne detector which measures the position quadrature of the pulse. The measurement result $X_j$ is stored. Each pulse entering the amplifier is subjected to a phase-space displacement along the position axis. The magnitude of the displacement applied to the $i$th pulse is proportional to \begin{equation}\label{dispamp}\sum_jJ_{ij}X_j,
\end{equation}
 where $X_j$ are the  results obtained from the measurements on other pulses. 

Thanks to this displacement, the sequence of phases obtained by the pulses at the end of the amplification cycle approaches the solution of the Ising problem, with the phases $\varphi_i=0$ and $\pi$ encoding the spin values of $\sigma_i=\pm1$, respectively. Intuitively, this can be understood as follows. Treating the spins as continuous variables and the Ising Hamiltonian as the potential energy, one finds the gradient of this energy with respect to $\sigma_i$ as $F=-\nabla_j H=\frac12\sum_j J_{ij}\sigma_j$. The physical interpretation of this gradient is the force that pushes the spin configuration towards lower energy. By applying that force in the form of displacement to the pulses circulating inside the OPO, we encourage their phases to align according to the minimum energy configuration.

\paragraph{CIM simulator.}
In SimCIM, we characterize each pulse by its complex amplitude $a_i = \frac{1}{\sqrt 2}(x_i + ip_i)$, where $x_i$ and $p_i$ are the canonical phase space quadratures. The per-roundtrip change in the amplitude can be written as 
\begin{equation}\label{DeltaPsi}
\Delta a_i=w a_i^*-\gamma a_i-s |a_i|^2a_i+\zeta \sum_jJ_{ij}x_j + f_i/\sqrt 2.
\end{equation}
In the above equation, the first term corresponds to the parametric gain, second to linear loss, third to nonlinear loss, fourth to the displacement and fifth to the noise. The nonlinear loss is associated with the second-harmonic generation, i.e. the transfer of energy from the signal back to the pump of the parametric amplifier. This effect becomes significant only for macroscopic signal amplitudes, but is necessary in the treatment in order to account for the saturation of the OPO. For the displacement term in \eeqref{DeltaPsi}, we are writing $\sum_jJ_{ij}x_j$ instead of \eqref{dispamp}  as if this term were calculated from the actual quadratures $x_j$ of the circulating pulses rather than the  quadratures $X_j$ measured on a tapped part of the optical energy. The additional quantum noises associated with this replacement are absorbed into the noise term $f_i$ together with the quantum noises arising from the losses.

We can now rewrite \eeqref{DeltaPsi} in terms of its real and imaginary components:
\begin{align}\label{DeltaXPi}
\Delta x_i&=w x_i-\gamma x_i-s (x_i^2+p_i^2)x_i+\zeta \sum_jJ_{ij}x_j + {\rm Re}f_i;\nonumber\\
\Delta p_i&=-w p_i-\gamma p_i-s (x_i^2+p_i^2)p_i + {\rm Im}f_i.
\end{align}
\begin{figure}
	\begin{tabular}{c}
\centerline{\includegraphics[width=0.5\linewidth]{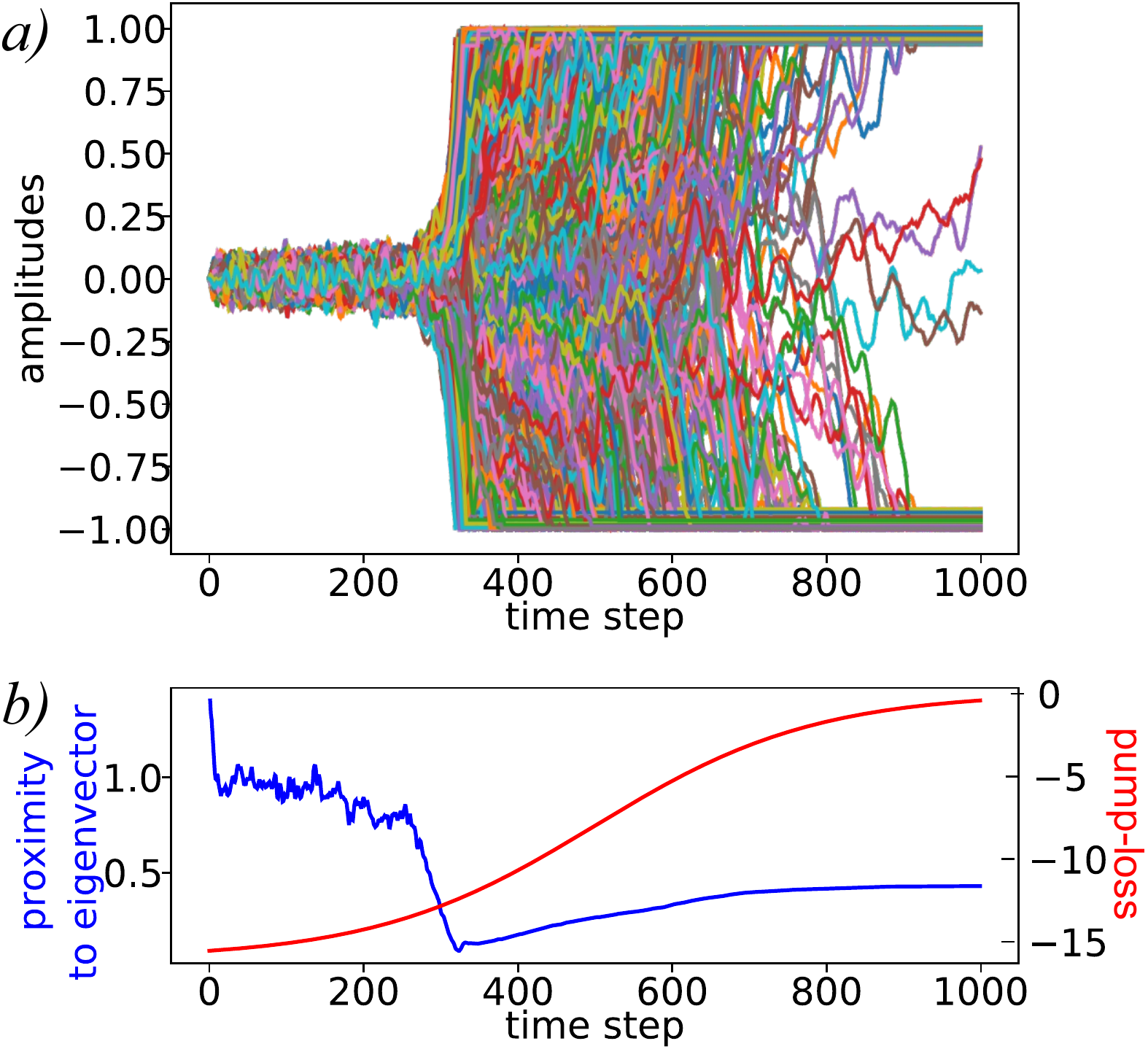}}
	\end{tabular}
	\caption{a) Evolution of the components of the ``spin" vector $\vec x=\{x_i\}$. b) Time dependence of the pump-loss factor $v$ and the quantity $||\mathcal N(\vec x) - \mathcal N(\hat J\vec x)||$ [the symbol $\mathcal N(\cdot)$ denoting normalization] which shows the proximity of $\vec x$ to the eigenvector of $\hat J$ with the highest eigenvalue.
	}\label{Prox_to_eig}
\end{figure}

In SimCIM, we omit the nonlinear loss term. This simplifies the calculation by allowing us to drop the second equation in \eqref{DeltaXPi} and restrict the analysis to real numbers: 
\begin{equation}\label{eq_alg}
\Delta{x}_i = v x_i + \zeta\sum_j J_{ij}x_j + f_i,
\end{equation}
where $v=w-\gamma$ represents both the parametric gain (controlled by the pump value) and the linear loss. To account for the saturation, we require that the amplitude $|x_i|$ not exceed a certain fixed value $x_{\rm sat}$. That is, we update the amplitude in each iteration according to $x_i\leftarrow\phi(x_i+\Delta x_i)$, where 
\begin{equation}\label{acti}
\phi(x)=\begin{cases}
x \textrm{ for } |x|\leq x_{\rm sat};\\
x_{\rm sat} \textrm{ otherwise}
\end{cases}
\end{equation}
is the ``activation function''. This approach is reminiscent to the Hopfield-Tank simulated annealer \cite{HopfieldTank}, but differs from it in that the activation function is applied directly to the ``neuron'' $x_i$ rather than its increment  $\Delta x_i$. A further difference is that this function is not differentiable. 

We implement this iterative procedure on the GeForce 1080 consumer videoprocessor utilizing the Tensorflow and PyTorch frameworks to take advantage of the parallel computation capability offered by this hardware. To accelerate the convergence, we use the momentum method \cite{Qian1999} with the momentum parameter of $\beta=0.9$.
We increase the pump-loss parameter $v$ according to the hyperbolic tangent law (Fig.~\ref{Prox_to_eig}). When it reaches a certain threshold level, the ``spins" $x_i$ start to grow and  eventually reach the limiting values of $\pm x_{\rm sat}$. While some spins reach these values quickly, others oscillate between positive and negative almost until the end of the run.

This dynamics can be understood if we rewrite \eeqref{eq_alg} in terms of the eigenvectors of the Ising matrix $\hat J$: 
\begin{equation}\label{eig}
\begin{cases}
\Delta{e_i} = (v+\zeta\Lambda_{ii}) e_i +f_i';\\
|\sum_j R_{ji}e_j|\le x_{\rm sat},
\end{cases}
\end{equation}
where $\vec e = \hat R\vec x$ and  $\hat\Lambda = \hat R \hat J \hat R^T$, with $\hat R$ being the orthogonal matrix that diagonalizes $\hat J$. In other words, if the pulse amplitudes comprise an eigenvector of $\hat J$, the evolution through the OPO will preserve that eigenvector as long as its amplitude is sufficiently small. In the initial stages of the evolution, the pumping is below the threshold; the ``spin" values fluctuate randomly around zero because of the noise term $f_i$. As the pump parameter $v$ ramps up, the value of $v+\zeta\Lambda_{ii}$ becomes positive for some $i$, leading to exponential growth of the corresponding eigenvector's amplitude. However, this growth will only continue until one of the eigenvector components reaches $x_{\rm sat}$, at which point the second part of \eeqref{eig} comes into play. After that, $\vec x$ is no longer an eigenvector of $\hat J$. As more and more components of $\vec x$ stabilize at $\pm x_{\rm sat}$, the effective matrix governing the dynamics of the remaining components changes, resulting in their seemingly chaotic oscillation between positive and negative values.

\paragraph{Results.}
\begin{figure*}
\includegraphics[width=\textwidth]{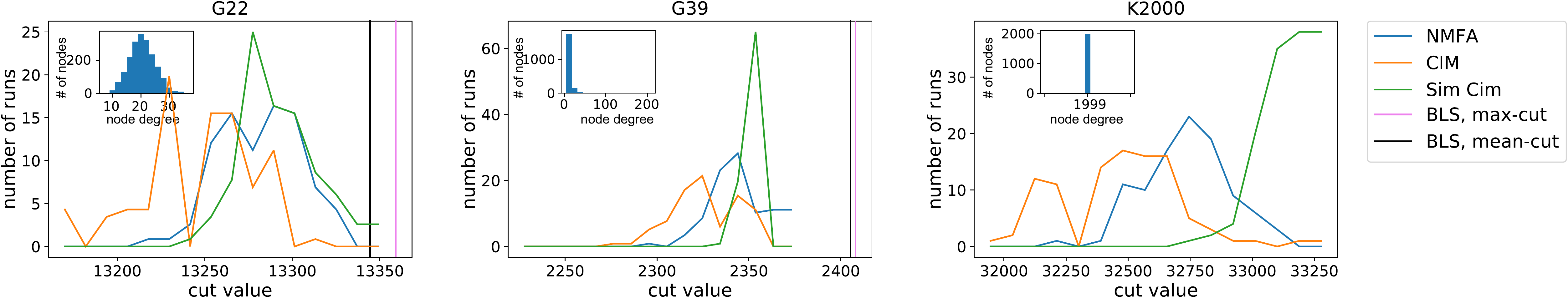}
\caption{Histograms for the graphs G22, G39 \cite{Gset} as well as the fully connected K2000 graph. For each graph the dependence between the number of runs and cut value is presented for the NMFA and SimCim algorithms and for CIM experimental results \cite{Mcmahon2016}. For G22 and G39 the results of CIM and algorithms also compared to BLS algorithm which gives the best known cuts.} \label{Res_G22_G39_K2000}
\end{figure*}

We present our simulation results in the same terms as in CIM papers \cite{Inagaki2016,Mcmahon2016}, namely the max-cut value problem \cite{Kar72}. In this problem, each edge of a certain graph has a value associated with it. The goal is to divide the nodes of a certain graph into two subsets such that the total value associated with the edges connecting them is maximized. This problem is equivalent to the Ising problem, as the cut value is given by 
\begin{equation}\label{cut}
{\rm cut} = -\frac{1}{2}\sum_{i<j}J_{ij}\left( 1 - \sigma_i \sigma_j \right), 
\end{equation}
where the spin value $\sigma_i = \pm 1$ defines which subset the $i$th node belongs to. The cut value is maximized if the Ising energy is minimized.

We compare our algorithm with the \emph{bona fide} CIM and the noisy mean field annealing (NMFA) algorithm \cite{King2018,MeanField}. In NMFA, the spin in each iteration is driven towards its self-consistent mean-field value $\tanh[\sum_jJ_{ij}x_j]$. To allow direct comparison with the CIM of Ref.~\cite{Inagaki2016}, we optimize each algorithm to run for 1000 iterations and run it 100 times, constructing the histogram of the cut value \eqref{cut}. 

We first run the simulations on the same set of 2000-node Ising graphs as Ref.~\cite{Inagaki2016}: relatively sparse graphs G22 and G39 from the G-Set \cite{Gset} and the fully connected graph K2000. The parameters of both SimCIM (the time-dependent gain-loss parameter $v$, overall feedforward factor $\zeta$ and the noise $f_i$) and NMFA have been optimized for each graph to obtain the best performance.  The results (Fig.~\ref{Res_G22_G39_K2000}) show superior performance of SimCIM with respect to the \emph{bona fide} CIM as well as visible improvement with respect to NMFA, with the advantage being particularly strong for the denser graph K2000. Additionally, we compare the performance with the known results of the breakout local search (BLS) \cite{BLS}, a classical algorithm which is known to perform well on the Ising annealing problem, albeit at a comparatively low speed. We see that the mean cut value obtained in the 100 SimCIM runs lies within 98\% of the best BLS result. 

A similar level of performance was observed on 800-node GSet graphs G1--G10. SimCIM with the same set of parameters reached the BLS ground state on 8 out of 10 graphs with the mean probability of $\sim25\%$. For the remaining two graphs (G2 and G9), the maximum cut value reached lay within 99\% of the BLS maximum. 

In terms of the computational speed, SimCIM is comparable to the CIM and NMFA. A single run of the CIM of Ref.~\cite{Inagaki2016} with 1000 roundtrips takes 5 ms. On the GeForce 1080 consumer videoprocessor, 100 runs of 1000-iteration SimCIM, launched in parallel, took 400 ms, corresponding to 4 ms per run. The corresponding figure for our implementation of NMFA was 5.5 ms. 

The \emph{bona fide} CIM must compute new displacement amplitudes \eqref{dispamp} at a rate that is faster than the pump pulse repetition rate (1 ns in Ref.~\cite{Inagaki2016}). In order to ensure that the FPGA keeps up with this rate, the designers of CIM restrict the allowed values of coupling terms in the Ising graph to $J_{ij}=\pm 1$ or $0$. Simulators, on the other hand, are not limited by this condition. 

\begin{figure}
\centerline{	\includegraphics[width=0.5\linewidth]{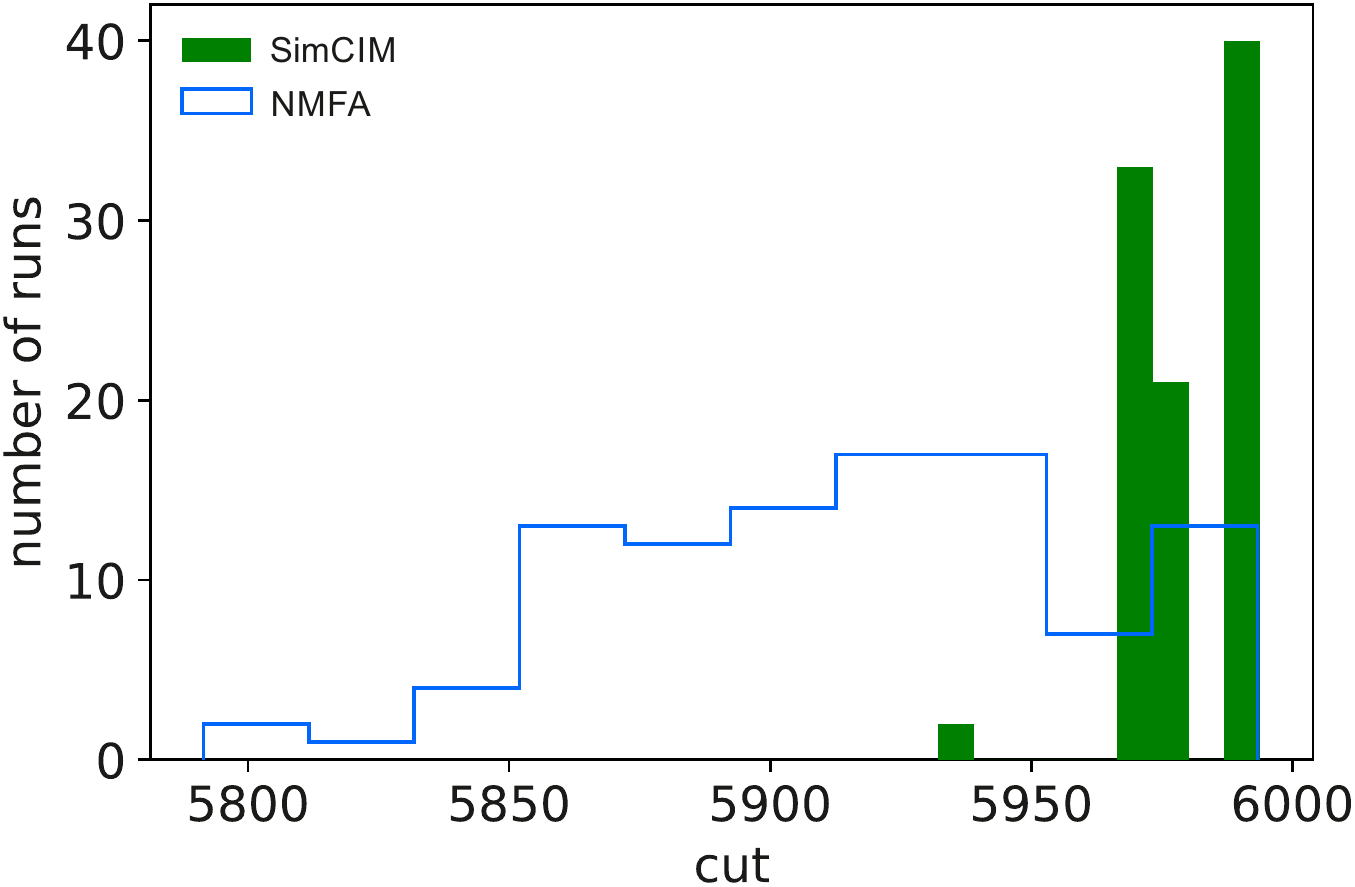}}
	\caption{Results for a random graph of  dimension 800 with real-valued couplings $J_{ij}$ that are normally distributed with zero mean and variance equal to one. }\label{Random_graph}
\end{figure}

To demonstrate this capability, we applied  NMFA and SimCIM to a set of 100 800-node graphs whose edge values are generated randomly according to a Gaussian distribution with the unit dispersion, running each algorithm 100 times for each graph. The same set of parameters has been used for the entire set. The results for one randomly chosen graph are shown in Fig.~\ref{Random_graph}. We found SimCIM to deliver consistently higher cut values, with the  cut value averaged over all runs and graphs equal to 5938 for NMFA and 6001 for SimCIM. If we consider the maximum cut value obtained by either algorithm in the 100 runs for each graph, averaged over all graphs, the two methods deliver similar values of 6022 for NMFA and 6021 for SimCIM.  The high quality of these results are not surprising, as it is known that the finding of the ground state is easier for all-to-all connected graphs than for sparse ones \cite{Dencegraphcomplexity}. Remarkably, however, SimCIM benefits from this to a larger extent than NMFA.
 
\paragraph{Discussion.}
Our results show that coherent Ising machines can be outperformed by a simulation algorithm running on a classical computer, with the added  advantage of the coupling values $J_{ij}$ not being limited to $0$ and $\pm1$. In this sense, our results contrast those of  Haribara {\it et al.} \cite{Haribara2017} who found that  classical annealing algorithms significantly underperform in comparison with the CIM in terms of the optimization time (this reference does not report any comparison in terms of the annealing quality). We believe this difference to be partially due to the hardware used (Ref.~\cite{Haribara2017} employed a CPU and a many-core processor whereas we used a GPU), and partially thanks to our algorithm specially designed to simulate and compete with the CIM.

There appears to be no added value either in the nonclassical nature of the optical states produced, or in the partially analog information processing  within the CIM. This is evident from the comparative performance analysis of CIM and SimCIM, and can be understood from Eqs.~\eqref{DeltaXPi} and \eqref{eq_alg} describing the simulator. The most computationally expensive part in each simulation step is the calculation of the feedforward term \eqref{dispamp}, which corresponds to the multiplication of a matrix and a vector. Both in SimCIM and the \emph{bona fide} CIM, this calculation is performed via a digital processor (GPU vs. FPGA, respectively). Hence it is not surprising that the analog optical processing in the CIM (which computes other terms in the simulation equation) does not bring about visible performance benefits. 

Asking ourselves, what is the feature responsible for the faster performance of the CIM compared to simulated annealing and taboo search algorithms, we are compelled to conclude that the CIM's primary asset is the different, albeit classical, underlying physics. 
A great potential of this quantum-inspired approach is warranted by the vast applicability of the NP-hard problem of annealing in the Ising system \cite{Smith1999}.

The potential applications of the Ising annealer can be classified into two groups, as summarized in a recent white paper by the D-Wave team \cite{DWaveWhitePaper}. First, a variety of optimization problems, such as the traveling salesman problem and generalizations thereof \cite{Neukart2017,Feld2018,Ohzeki2018}, financial portfolio optimization \cite{Rosenberg2016}, protein folding \cite{Perdomo2012}, as well as constraint satisfaction problems, e.g. factorization \cite{Factoring_Guzik} and satisfiability \cite{Douglass2015,Bian2018}, can be reduced to the Ising optimization. Second, sampling of low-lying energy states is a useful tool for machine learning, specifically for the training of Boltzmann machines, which are in turn the primary component of deep belief neural networks. A related application of the SimCIM and the restricted Boltzmann machine based thereupon is the simulation of quantum condensed matter systems and phase transitions therein \cite{Carleo2017}, as well as quantum state and process tomography \cite{Torlai2018}. 

As mentioned previously, CIM's alleged superiority in comparison with the D-Wave annealer \cite{CIMvsDwave} has been challenged by McGeogh {\it et al.} \cite{CIMvsDwaveComment}. This paper criticizes the performance metric used in Ref.~\cite{CIMvsDwave} and furthermore argues that the comparison was made on Ising graphs that are known to be easy to solve for classical machines. This criticism has, in turn, been responded to in the second version of Ref.~\cite{CIMvsDwave}. In summary, it remains an open question on which of the above problems our simulator will prove competitive with respect to both classical heuristics and NISQ devices.

Further improvement to the SimCIM can be obtained by using it in combination with classical taboo search algorithms \cite{Blanzieri2018} and by 
on-the-fly optimization of the simulation parameters based on the real time observation of individual spin trajectories and the features of the Ising graph. Leleu {\it et al.} used such an approach to eliminate stable equilibria associated with the local minima of the Ising Hamiltonian and therefore enhance the likelihood of reaching the global minimum. For some GSet graphs, they demonstrated cut values above those obtained by BLS \cite{leleu2018destabilization}. 

Another interesting potential research direction would be the experimental implementation of the CIM-like machine so that the feedforward term is not calculated digitally but found through optical interference of individual modes (which would need to be synchronized in time). This can be implemented in either a free-space \cite{Marandi2014} or integrated \cite{Roques2018} fashion. The computational speedup would then be obtained thanks to the natural parallelism of analog optical processing. An important observation we make from Fig.~\ref{Prox_to_eig} is that the gain-loss parameter can be kept negative throughout the simulator run. This means that the parametric gain element may in fact not be necessary in the optical implementation of the CIM (or its successors) provided that the per-roundtrip loss is maintained sufficiently small.  

We thank Alireza Marandi for stimulating discussions, Kirill Kalinin for help in calculations and Natalia Berloff for making us aware of Hopfield networks. The work of A. L. is supported by the RFBR (Grant No. 18-37-20033).


\end{document}